\documentclass[iop]{emulateapj}

\usepackage{epstopdf}
\usepackage{epsf}
\usepackage{graphics}
\usepackage{natbib}

\usepackage{graphicx}
\usepackage{bm}
\usepackage{natbib}
\usepackage{amssymb}

\newcommand{\msun}{\mbox{$M_{\odot}$}}

\newcommand{\lsun}{\mbox{$L_{\odot}$}~}
\newcommand{\kms}{\mbox{km s$^{-1}$}}

\newcommand{\etal}[1]{{ et al.}~}
\def\kms{\ifmmode \hbox{km~s}^{-1}\else km~s$^{-1}$\fi}
\def\etal {{\it et al.}}
\def\deg      {{\ifmmode^\circ\else$^\circ$\fi} } 

\def\h2     {H$_2$}
\def\arcdeg{\hbox{$^\circ$}}

\def\arcsec{\hbox{$^{\prime\prime}$}}

\shorttitle{Submm Recombination Lines}
\shortauthors{Scoville \& Murchikova}

\date{}                                           

\begin{document}

\title{Submm Recombination Lines in Dust-Obscured Starbursts and AGN}
 \author{ N. Scoville\altaffilmark{1} and L. Murchikova\altaffilmark{2}}

\altaffiltext{1}{California Institute of Technology, MC 249-17, 1200 East California Boulevard, Pasadena, CA 91125}
\altaffiltext{2}{California Institute of Technology, MC 350-17, 1200 East California Boulevard, Pasadena, CA 91125}



\altaffiltext{}{}

\begin{abstract}
We examine the use of submm recombination lines of H, He and He$^+$ to probe the extreme ultraviolet (EUV) luminosity of 
starbursts (SB) and AGN. We find that the submm recombination 
lines of H, He and He$^+$ are in fact extremely reliable and quantitative probes of the EUV continuum at 
13.6 eV to above 54.6 eV. At submm wavelengths, the recombination lines originate from low energy levels (n = 20 -- 50). The maser amplification, 
which poses significant problems for quantitative interpretation of the higher n, radio frequency recombination lines, is insignificant.
Lastly, at submm wavelengths the dust extinction is minimal. The submm line luminosities are therefore directly proportional to 
the emission measures ($EM_{ION} = n_e \times n_{ion} \times \rm{volume}$) of their 
ionized regions. We also find that the 
expected line fluxes are detectable with ALMA and can be imaged 
at $\sim0.1$\arcsec ~resolution in low redshift ULIRGs. Imaging of the HI lines will provide accurate spatial and kinematic mapping of the star formation distribution in low-z IR-luminous galaxies. And the relative fluxes of the HI and HeII recombination lines will strongly constrain the relative contributions of 
starbursts and AGN to the luminosity. The HI lines should also provide an avenue to constraining the submm dust extinction curve.

\end{abstract}

\medskip


\section{Introduction}
The most energetic periods of evolution in galaxies are often highly obscured by dust at short 
wavelengths, with the luminosity reradiated in the far infrared. Merging of galaxies will concentrate 
the interstellar gas and dust (ISM) in the nucleus  since the gas is very dissipative where it 
can fuel a nuclear starburst or AGN. The Ultraluminous Infrared Galaxies (ULIRGs) and Submm 
Galaxies (SMGs) emit nearly all their radiation in the far infrared \citep{san88,car13}. Although their power originates 
as visible, UV and X-ray photons, the emergent IR continuum only weakly differentiates the power source(s)
-- starburst or AGN -- and their relative contributions. This is a significant obstacle to understanding the
evolution of the nuclear activity since the star formation and AGN fueling may occur at different stages 
and with varying rates for each. Many of the signatures of star formation or black hole activity (e.g. 
X-ray, radio or optical emission lines) can be indicative that starbursts or AGN are present but 
provide little quantitative assessment of their relative contributions or importance \citep[a summary of the various SFR indicators is provided in][]{mur11}. 

In this paper we develop the theoretical basis for using the submm recombination lines of H, He and He$^+$ 
to probe star formation and AGN. We find that the emissivities of these lines can provide reliable 
estimates of the EUV luminosities from 13.6 eV to $\sim$$10^2$ eV and hence the relative luminosities 
associated with star formation (EUV near the Lyman limit) and AGN accretion (harder EUV). 

Although the extremely high infrared luminosities of ULIRGs like Arp 220 and Mrk231 are believed to be powered by 
starburst and AGN activity, the distribution 
of star formation and the relative contributions of AGN accretion is very poorly constrained. This is due to inadequate angular resolution in the 
infrared and the enormous and spatially variable extinctions in the visible ($A_{V} \sim 500 - 2000$ mag). The submm lines
will have minimal dust extinction attenuation. 
And, given the large number of recombination lines across the submm band, lines of the different species 
may be found which are close in wavelength and provide the capability to move to longer wavelengths to further reduce the dust opacity (in the 
most opaque sources). Although mid-IR fine structure transitions of heavy ions have been used 
in some heavily obscured galaxies, the line ratios depend on density, temperature and metallicity;
in contrast, the H and He$^+$ lines have none of these complications. Lastly, we find that the expected 
fluxes in the lines are quite readily detectable with ALMA.

In the following, we first derive the emissivities and line opacities for the submm recombination lines as a function of
density and temperature (\S \ref{emiss}). Then using simplified models for the ionizing continuum associated 
with OB stars and with AGN, we derive the relative emission measures of the H$^+$, He$^+$ and He$^{++}$ 
regions for these two EUV radiation fields (\S \ref{ioniz}). Lastly, we compare the expected line fluxes in HI and HeII 
with the sensitivity of ALMA and find that the lines should be readily detectable from ULIRG  
nuclei at low z. The observations of these lines can therefore provide the first truly quantitative assessment 
of the relative contributions of starbursts and AGN to the luminosity of individual objects.

\section{Submm Recombination Lines}\label{emiss}

The low-n HI recombination lines at mm/submm wavelengths trace the emission-measure of the ionized gas and 
hence the Lyman continuum production rate associated with high mass stars and AGN. 
In contrast to the m/cm-wave radio HI recombination lines which can have substantial maser amplification \citep{bro78, gor90, pux97}, the submm recombination emission is predominantly 
spontaneous emission with relatively little stimulated emission and associated non-linear amplification (see \S \ref{maser}).
Since the submm HI lines (and the free-free continuum) are also  
optically thin, their line fluxes are a linear tracer of the ionized gas emission measure (EM = n$_e$ n$_p$ vol). 
Hence these lines are an excellent probe of the EUV luminosity
of OB stars and AGN (assuming the EUV photons are not appreciably absorbed by dust). Lastly, we note that 
in virtually all sources, the dust extinction of the recombination lines at $\lambda \sim 350\mu$m to 1mm will be insignificant. 


Early observations of the mm-wave recombination lines were made in Galactic compact HII -- in these regions the continuum 
is entirely free-free and hence one expects fairly constant line-to-continuum flux ratio if the mm-line emission arises from spontaneous 
decay in high density gas with little stimulated emission contribution.  This is indeed the case -- \cite{gor90} observed 
the H40$\alpha$ line at 99 GHz  in 7 HII regions and found a mean ratio for the integrated-line brightness (in K  km s$^{-1}$) to continuum   
of 31.6  (K km s$^{-1}$/K). Less than 3\% variation in the ratio is seen across the sample. The optically thin free-free emission provides a 
linear probe of the HII region emission measure (EM) and hence the OB star Lyman continuum production rate.  The observed constancy of the line-to-continuum ratios then 
strongly supports the assertion that the integrated recombination line fluxes are also a linear probe of the Lyman continuum production rates.

\subsection{HII Line Emissivities}\label{h_emiss}

\begin{figure}
\epsscale{1.0}  
\plotone{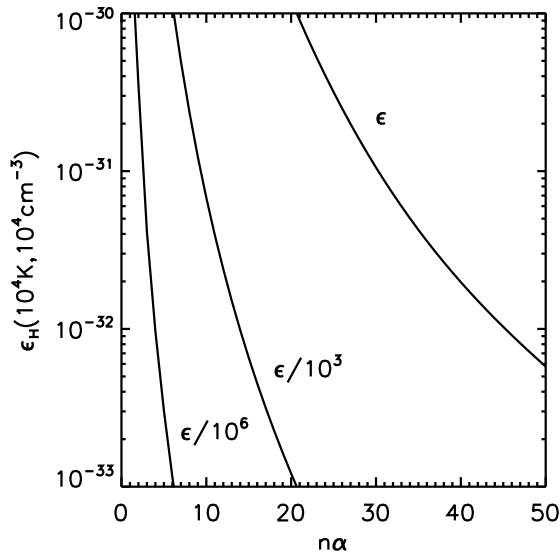}
\caption{ The emissivities of the HI $\alpha$ recombination lines are shown for T = 10$^4$ K. The emissivity 
is given per unit n$_e$ n$_p$ and the x-axis has the lower quantum number of the H$\alpha$ transitions.  
The emissivities were calculated for n$_e = 10, 1000,$ and $10^4$ cm$^{-3}$; and the curves for all densities are coincident.  }
\label{hi_emiss} 
\end{figure}

\begin{figure}
\epsscale{1.0}  
\plotone{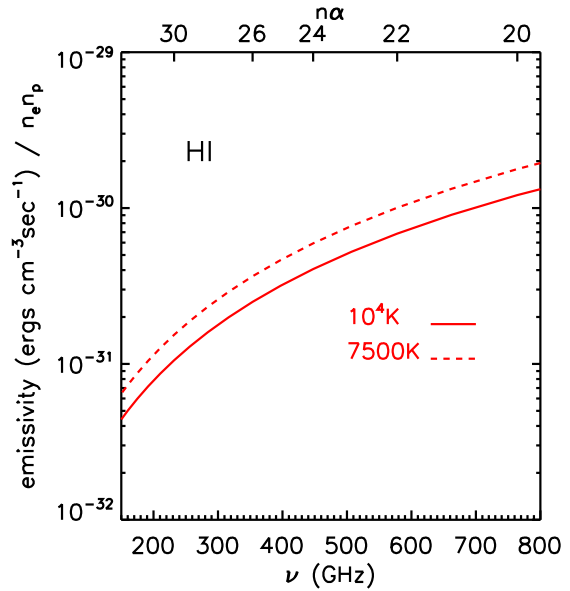}
\caption{ The emissivities of the HI submm recombination lines are shown for T = 7500 and 10$^4$ K. The emissivity 
is given per unit n$_e$ n$_p$. The top border has the quantum number of the $\alpha$ transitions and their rest frequencies are on the bottom axis.  
The emissivities were calculated for n$_e = 10, 1000,$ and $10^4$ cm$^{-3}$; and the curves for all densities are coincident.  }
\label{hi_emiss_submm} 
\end{figure}

To calculate the expected  HI line emission we make use of standard recombination line analysis  \citep[as described in][]{ost06}.
The volume emissivity, $\epsilon$ is then given by 
\begin{eqnarray}
 \epsilon &=& n_u A_{ul} h\nu  \nonumber \\
 &=& b_{n_{u}} n_u(TE) A_{ul} h \nu
\end{eqnarray}
\noindent where $n_u$ and $n_u(TE)$ are the actual and thermal equilibrium upper-level population densities.  The exact HI spontaneous decay rates from levels u to l, A$_{ul}$, are available in tabular form online from \cite{kho05}. The most complete and up to date departure coefficients (from TE)(b$_n$ and $d (\ln b_n) / dn$ are from \cite{hum87,sto95,sto95a}. The 
latter work includes population transfer by electron and ion collisions and has emissivities for HI and HeII up to principal quantum number n = 50. They also calculate optical depth parameters for a wide range of electron temperature (T$_e$) and electron density ($n_e$). We make use of these numerical results in this paper; Figure \ref{hi_emiss} shows the \cite{sto95a} HI recombination line emissivities at T = $10^4$K. 

Figure \ref{hi_emiss_submm} shows the submm HI-n$\alpha$ line emissivities $\epsilon$ per unit emission measure,  for T = 7500 and 10$^4$ K. These volume emissivities were computed 
for density n = 10$^2$ and 10$^4$ cm$^{-3}$ but the separate density curves are essentially identical. This is because, for these low energy levels, the spontaneous decay rates are very high (A$_{\Delta n = 1} (HI) > 200$ sec$^{-1}$ for n $< 30$). The level populations are therefore determined mainly by the radiative cascade following recombination to high levels. The latter 
is proportional to the recombination rate and hence $n_e n_p$.

To translate the curves in Figure \ref{hi_emiss_submm} into expected emission line 
luminosities, one needs to multiply by the total emission measure of each source. Consider the detectability of a luminous star forming region in a nearby galaxy. In \S \ref{ioniz} we show that for a starburst type EUV 
spectrum with integrated luminosity in the ionizing continuum at $\lambda < 912$\AA, L$_{EUV} = 10^{12}$ \lsun, the total Lyman continuum photon production rate is $Q_{LyC}= 1.20\times10^{56}$ sec$^{-1}$. 
Scaling this down to the luminosity of an OB star cluster with L$_{EUV} = 10^{6}$ \lsun gives $Q_{LyC}= 1.20\times10^{50}$ sec$^{-1}$.  
For Case B recombination in which all the photons are absorbed (i.e the HII is ionization bounded) and the Lyman 
series lines are optically thick, the Lyman lines above Ly$\alpha$ don't escape. (In fact, most of the Ly$\alpha$ may be absorbed by any residual dust.)
In this case, the standard Str\"{o}mgren condition equating the supply of fresh Lyman continuum photons (Q$_{Lyc}$) to the volume integrated rate of recombination to states above the 
ground state, 
\begin{eqnarray}
Q_{LyC}  &=& \alpha_B n_e n_p vol  ,
\end{eqnarray}
\noindent implies an HII region emission measure (EM = $n_e n_p$ vol) 
of EM = $4.60\times10^{62}$ cm$^{-3}$ (using $\alpha_B = 2.6\times10^{-13}$ cm$^{3}$ sec$^{-1}$ at $T_e = 10^4$ K). Using the specific emissivity of $3\times10^{-31}$ ergs cm${-3}$ sec$^{-1}$ for HI-26$\alpha$ from Figure \ref{hi_emiss_submm}, the recombination line luminosity will be L$_{\rm H26\alpha} = 1.38\times10^{32}$ ergs sec$^{-1}$. For a source distance of
1 Mpc and a line width of 30 \kms, this corresponds to a peak line flux density of $\sim 3.3$ mJy.  This flux density is readily detectable at signal to noise ratio 10$\sigma$ within $\sim1$ hr with ALMA Cycle 1 sensitivity. 

\subsection{He Line Emission}\label{he_emiss}

HeI has an ionization potential of 24.6 eV and its photoionization requirements are not very different 
than those of HI. Thus the HeI recombination lines probe the ionizing UV radiation field in much the same way 
as HI (see \S \ref{ioniz}). Since the HeI submm lines will be weaker than those of HI due to the lower He abundance, we don't examine the HeI emission extensively here and instead focus on HeII.

\begin{figure}
\epsscale{1.0}  
\plotone{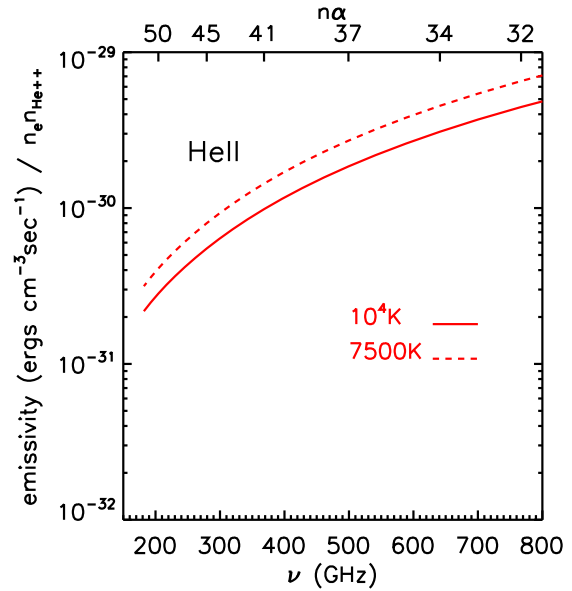}
\caption{ The emissivities of the HeII submm recombination lines are shown for T = 7500 and 10$^4$ K. For HeII it is per unit n$_e$ n$_{he++}$.The top border has the quantum number of the $\alpha$ transitions and their rest frequencies are on the bottom axis.  
The emissivities were calculated for n$_e = 10, 1000,$ and $10^4$ cm$^{-3}$; and the curves for all densities are coincident.  Note that the emissivities for HeII
per unit n$_e$ n$_{He++}$ are 4-5 times greater than those for HI at similar frequency -- this partially compensates for the lower He abundance relative to 
H if their ionized volumes are similar (as would be the case for a very hard ionizing continuum).}
\label{he_emiss_submm} 
\end{figure}

The ionization potential of HeII is 54.4 eV corresponding to photons with $\lambda = 228$ \AA~ for conversion of He$^+$ to He$^{++}$. 
Since the most massive star in a  starburst will have surface temperatures $\sim 50000$ K, the ionizing EUV from such a population 
will have only a very small fraction of the photons with energy sufficient to produce He$^{++}$. Thus the recombination lines of HeII (He$^{+}$) 
which are produced by recombination of e + He$^{++}$ can be a strong discriminant for the existence of an AGN with a relatively hard 
EUV-X-ray continuum. In starbursts there can be some HeII emission associated with Wolf-Rayet stars. However, the emission measure of the He$^{++}$ region relative to that of the H$^+$ region will be much less than for an AGN.

The emissivities of the HeII recombination lines are taken from \cite{sto95,sto95a}. For the interested reader, a simple model for the scaling of rate coefficients between Hydrogen and hydrogenic ions is developed analytically in Appendix \ref{appen_scale}
and those relations are compared with the numerical results from \cite{sto95a} in Appendix \ref{appen_comp}.  

The HeII 
submm $\alpha$ lines are at higher quantum numbers n than those of HI since the energy levels scale as the nuclear charge Z$^2$, i.e. a factor of 4 larger for the same principal quantum number n in He$^+$. For the submm HeII transitions, n = 30 - 50, versus 20 - 35 for HI.  In Figure \ref{he_emiss_submm} the expected HeII line emissivities per unit n$_e$ n$_{He++}$ are shown for the submm band. The values of these emissivities are
$\sim5$ times those of HI (Figure \ref{hi_emiss}; however, since the He/H abundance ratio is 0.1 the actual values per unit n$_e$ n$_{p}$ are quite similar in a plasma where all the H is ionized and all the He is He$^{++}$.

\subsection{HeII/HI Emission Line Ratios with T$_e$ and n$_e$}\label{he_emiss_ratios}

\begin{figure}
\epsscale{1}  
\plottwo{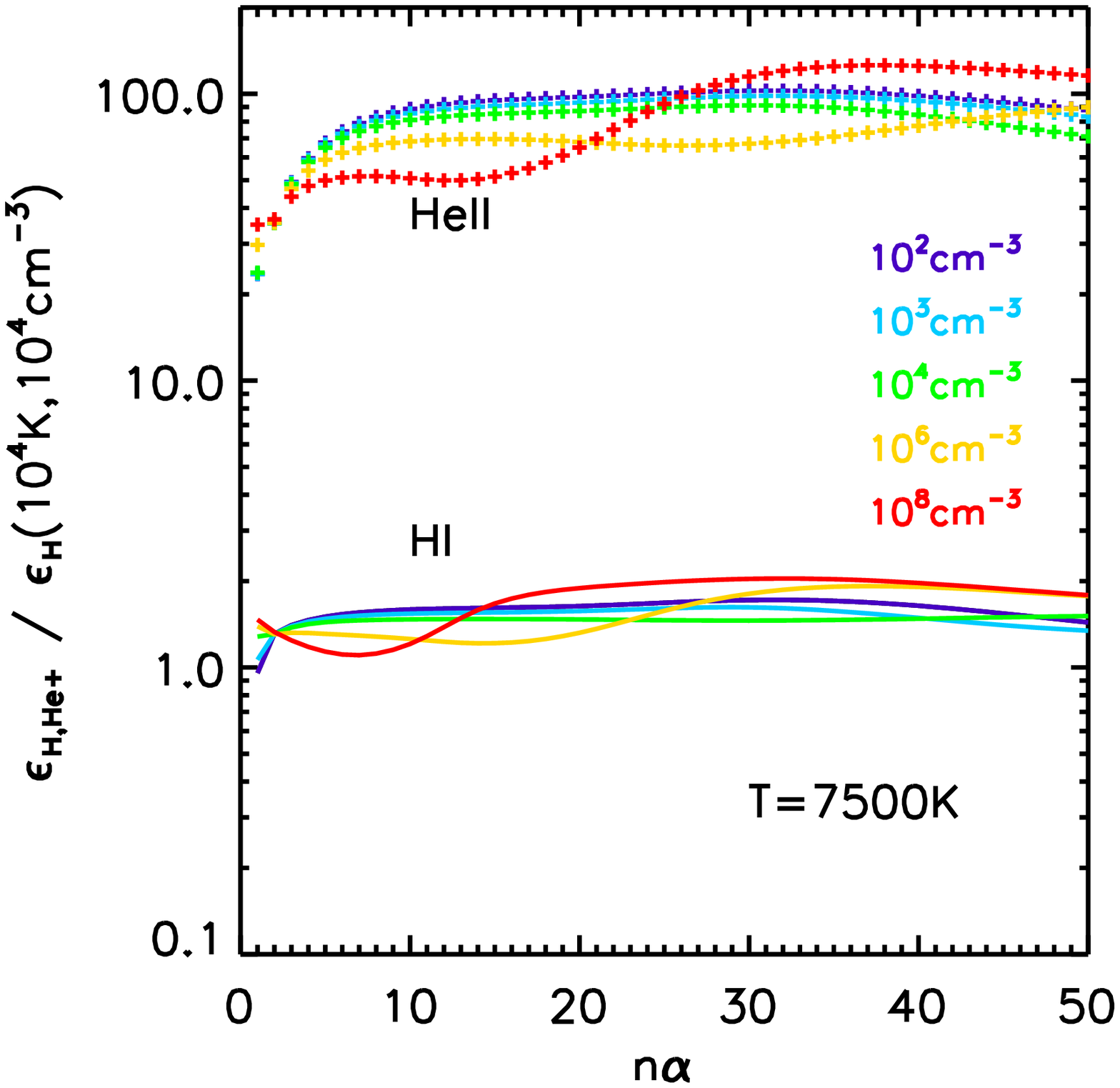}{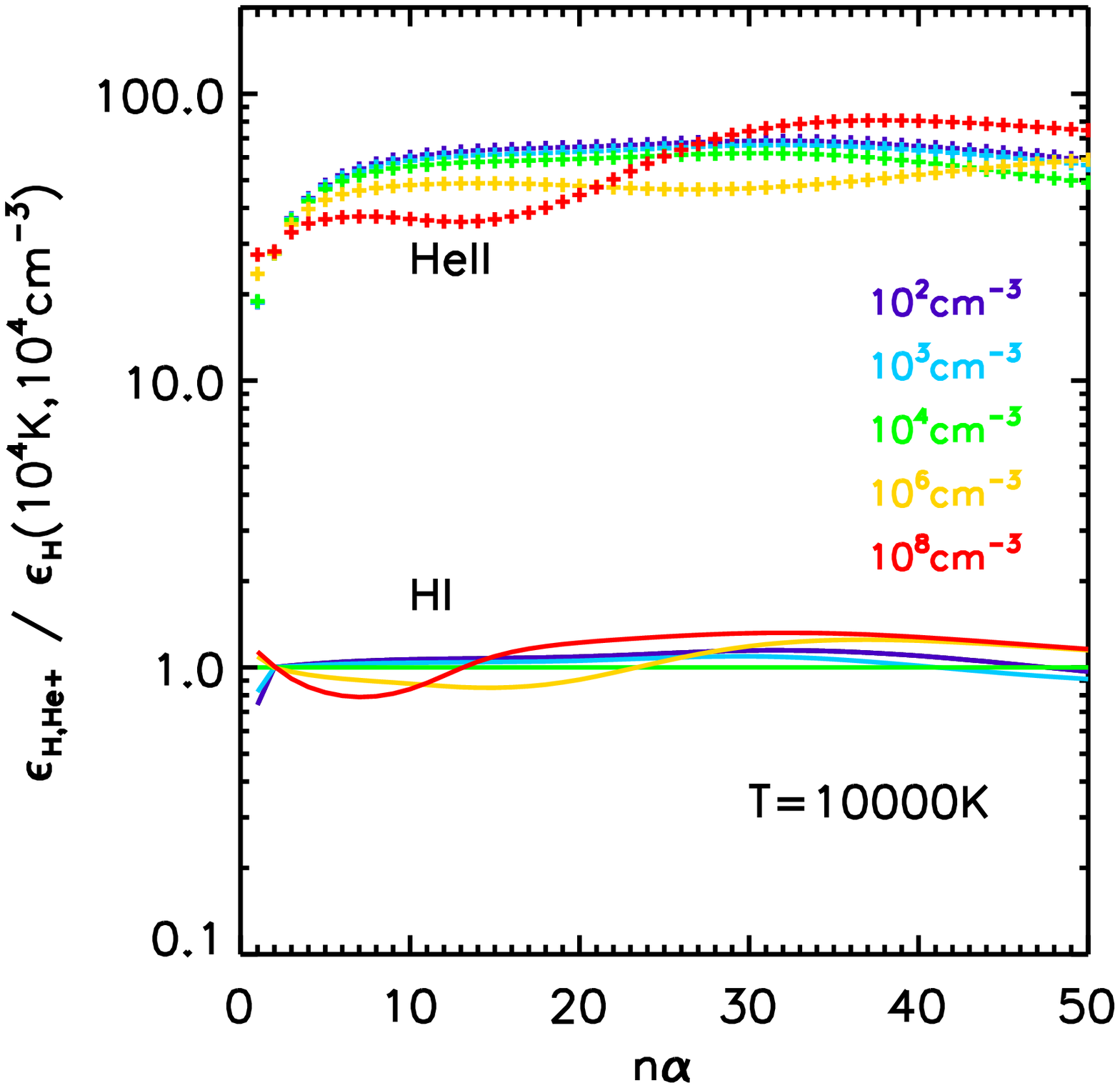}
\plottwo{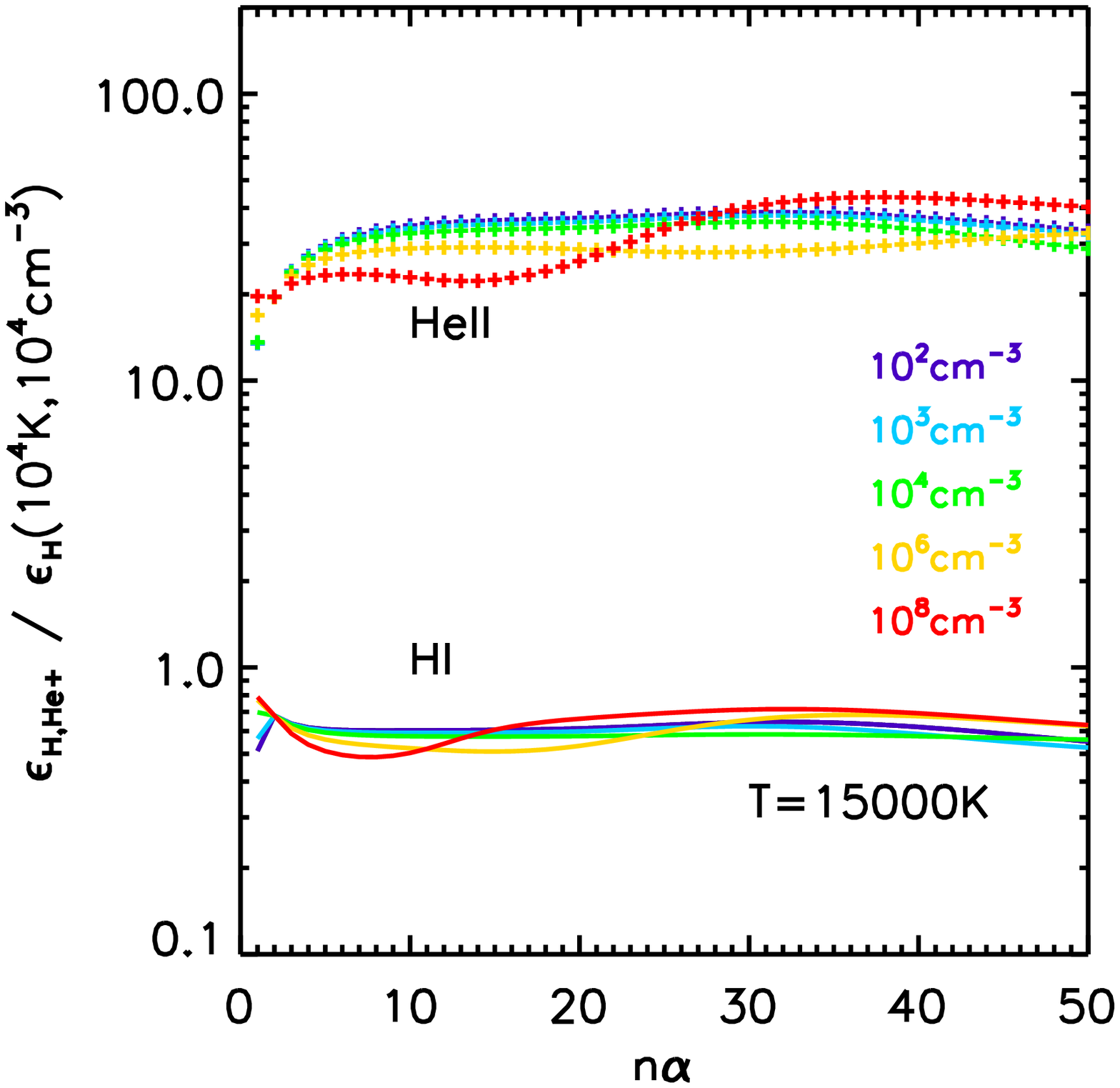}{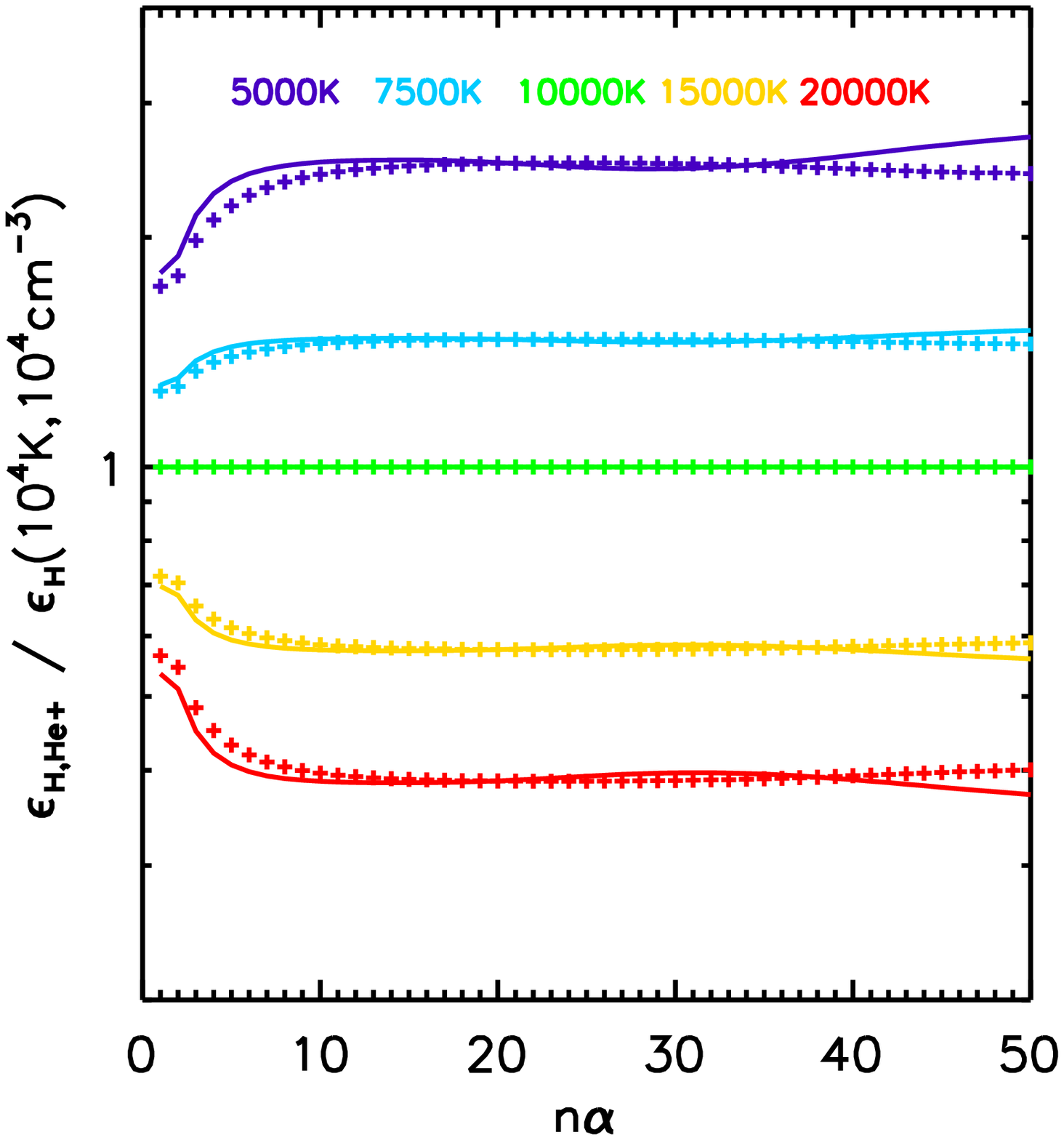}
\caption{The emissitvity ratios of HeII to HI are shown as a function of principal quantum number n for n$_e = 10^2 - 10^8$ and T$_e$ = 7500, 10000 and 15000 K. [The emissivities are normalized to HI at $10^4$ K and density $10^4$ cm$^{-3}$ so one can see the dependence on ion, temperature and density. 
The line ratios are very nearly independent of density for all the temperatures but they do depend on temperature as $T_e^{-4/3}$ for quantum number n $\sim$ 10 to 50.. The latter 
is clearly shown in the lower-right panel where the emissivity ratio is shown for 5 temperatures at n$_e = 10^4$ cm$^{-3}$.   }
\label{he_rel_h} 
\end{figure}

In Figure \ref{he_rel_h} 
the ratios of HeII/HI $\alpha$ recombination line emissivities \citep{sto95} are shown as a function of principal quantum number for large ranges of both 
T$_e$ and n$_e$.  Two cautions in viewing these 
plots: 1) as noted above the lines of HI and HeII are not at the same frequency for each n and 2) the emissivity ratios are per n$_e$ n$_{He{}^{++}}$ and per n$_e$ n$_{p}$ for HeII 
and HI, respectively. In the case of the latter, the EM for He$++$ will be almost always $< 0.1$ of that for H$+$ due to the lower cosmic abundance of He. 

 Figures \ref{he_rel_h}-top and lower-left clearly show that at a given temperature, T$_e$, the HeII/HI line ratio is virtually constant as a function
 of both quantum number and electron density. Thus, varying density in the ionize gas should have almost no influence on the line ratios of HeII to HI. 
 On the other hand, it is clear from these figures that increasing T$_e$ leads to a decrease in the HeII/HI emissivity ratio. In Figure \ref{he_rel_h}-lower-right, 
 the ratio is shown for a single density $n_e = 10^4$ cm$^{-3}$ but T$_e = 5000$ to 20,000 K and the temperature dependence is clear and 
 the same for all n transitions. Thus, the temperature and density dependence of the HeII to HI line ratios at fixed n$\alpha$ can be empirically fit by:
 
\begin{eqnarray}
{\epsilon_{HeII-n\alpha} \over{ \epsilon_{HI-n\alpha}}} &\propto& n_e^0 ~T_e^{-4/3}  .
\end{eqnarray}

\noindent Although the HeII/HI line ratios are temperature dependent, the actual range of temperatures expected for the ionized gas is very
limited, T$_e =7500 - 10000$ K in star forming HII regions due to the strong thermostating of the cooling function which decreases strongly at lower 
temperatures and increases steeply at higher temperatures \citep[see][]{ost06}. For the AGN sources, it is also unlikely that the temperatures will be much
higher since most of the heating is still provided by Lyc photons near the Lyman limit (even though there are harder photons in their EUV spectra). 

In Appendix \ref{appen_scale}, we show that the recombination 
rates coefficients scale as :
\begin{eqnarray}
\alpha_{He+} (T_e) = Z  \alpha_{H} (T_e / Z^2)  ~{\rm with ~Z~=~2~for~He}^+.
\end{eqnarray} 
\noindent However, the line emission rates also depends on the radiative and collisional cascade through the high n levels and it is not possible to derive the emissivity scaling analytically to better than a factor of 2 accuracy.

\subsection{Maser Amplification ?}\label{maser}

\begin{figure}
\epsscale{1.0}  
\plottwo{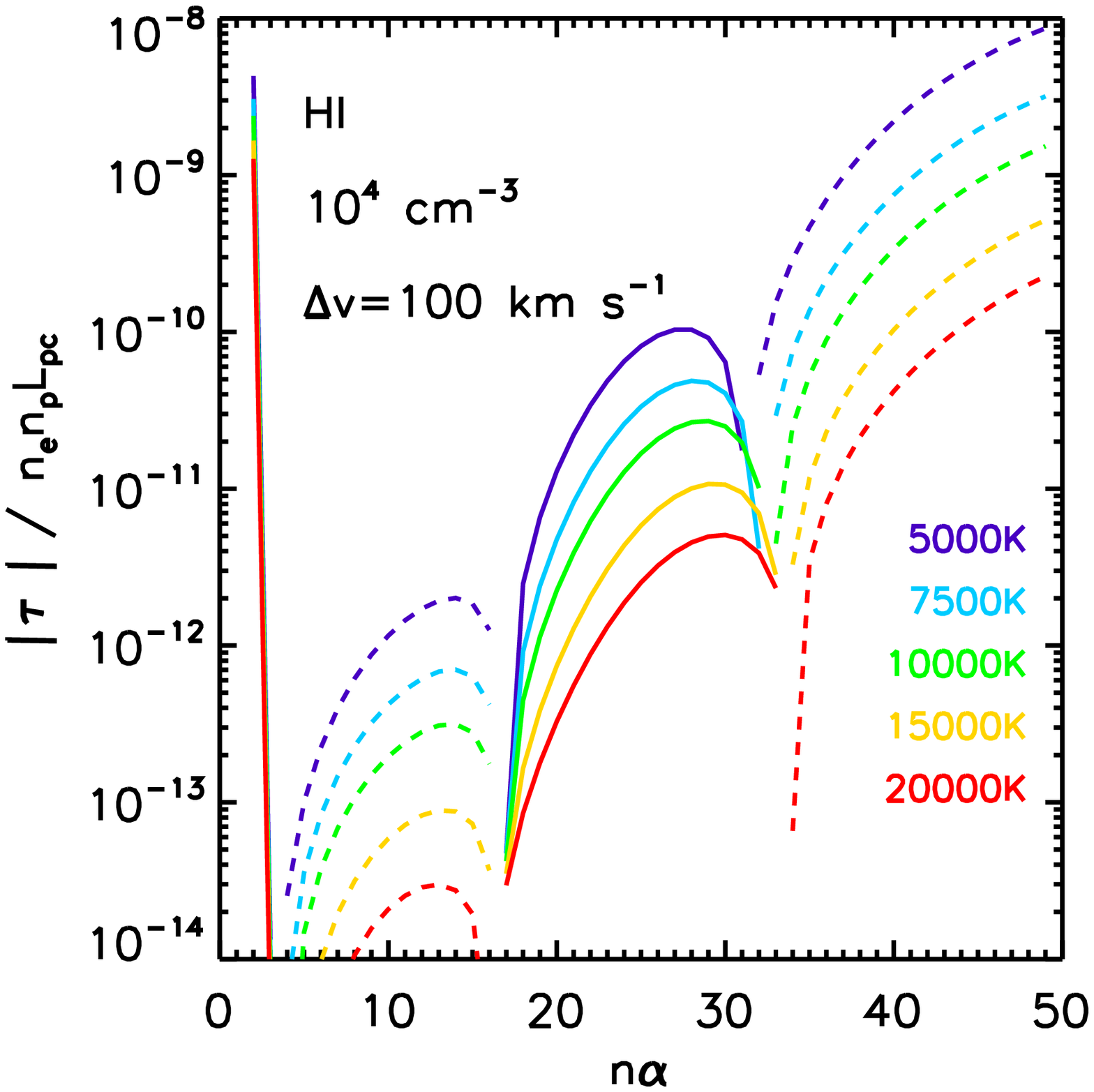}{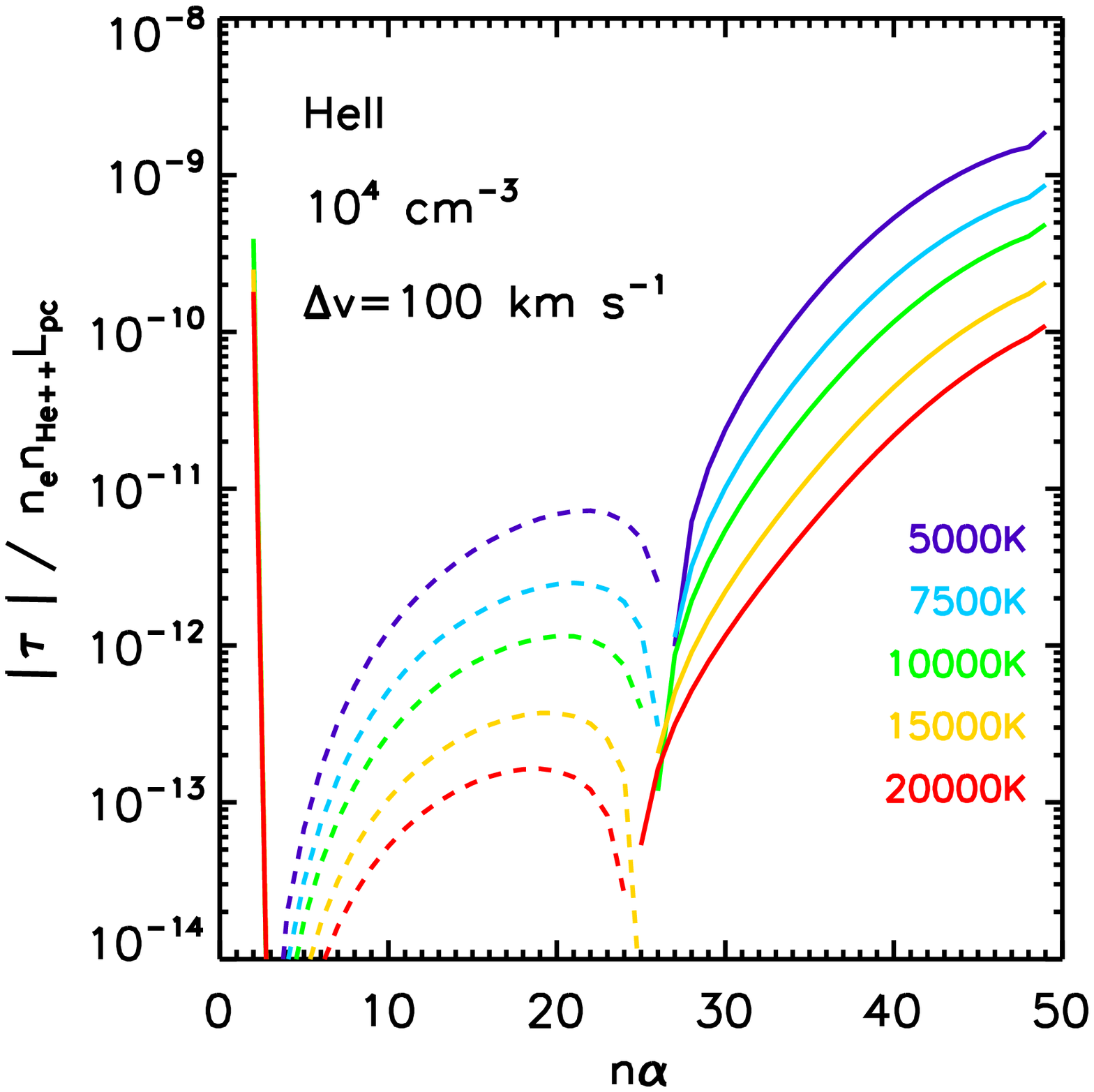}
\caption{ The {\it specific} optical depths of the HI and HeII recombination lines are shown with level populations calculated  for T = 5000 to 20,000 K and density n$_e$ = $10^4$ cm$^{-3}$. 
A fiducial line width of 100 \kms was adopted. The specific optical
depth is per unit $n_e n_p L_{pc}$ and the lines are dashed where the optical depth is negative, implying a population inversion and possible 
maser amplification. A likely maximum line-of-sight emission measure is $\int n_e n_p dl \sim 10^9$ cm$^{-6}$ pc, corresponding to a ULIRG  
starburst nucleus.  
}
\label{hi_tau1} 
\end{figure}

\begin{figure}
\epsscale{1.0}  
\plottwo{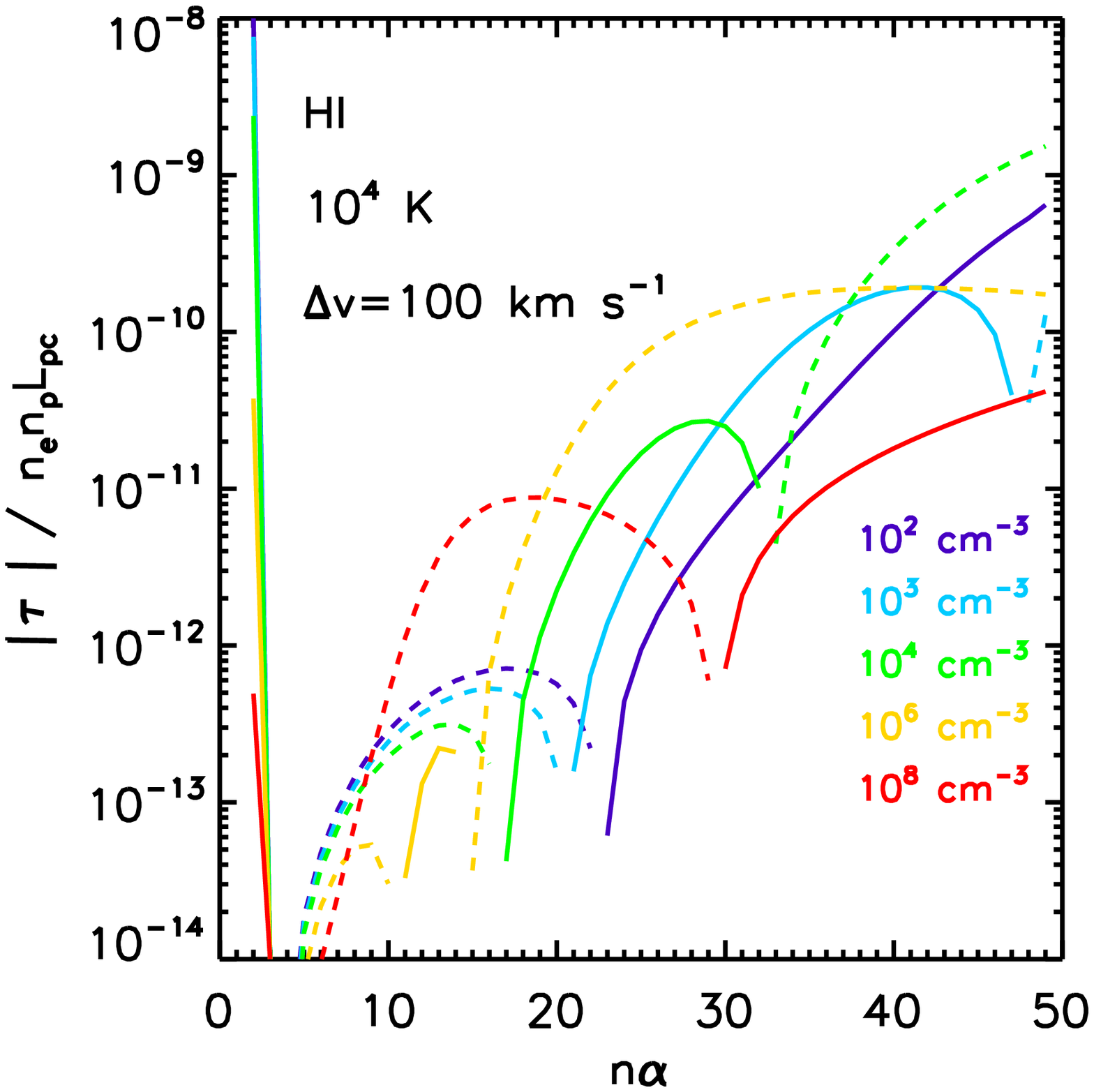}{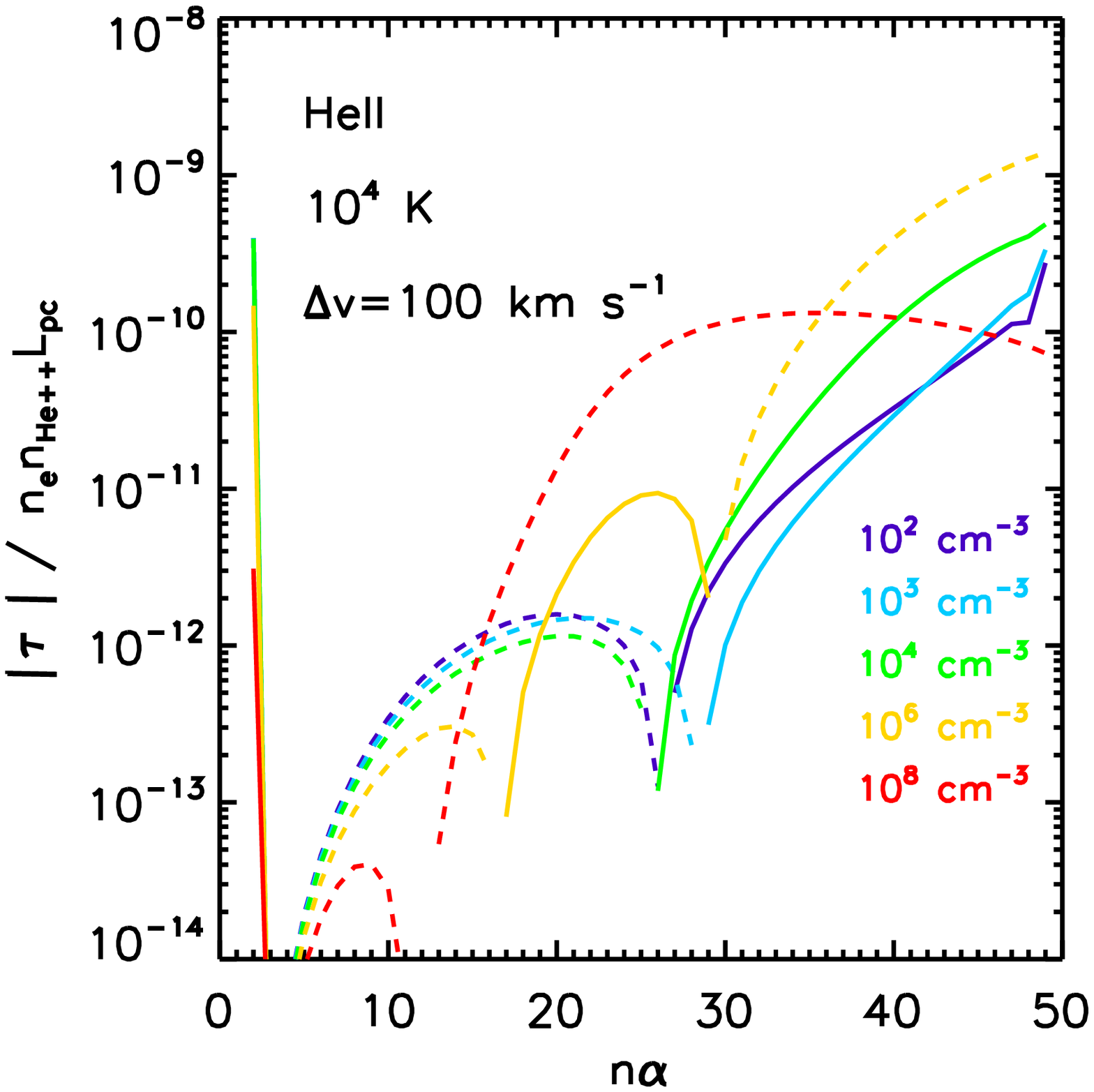}
\caption{ The {\it specific} optical depths of the HI and HeII recombination lines are shown with level populations calculated for T = $10^4$ K and densities ranging between n$_e$ = $10^2$ and $10^8$ cm$^{-3}$. 
A fiducial line width of 100 \kms was adopted. The specific optical
depth is per unit $n_e n_p L_{pc}$ and the lines are dashed where the optical depth is negative, implying a population inversion and possible 
maser amplification. A likely maximum line-of-sight emission measure is $\int n_e n_p dl \sim 10^9$ cm$^{-6}$ pc, corresponding to a ULIRG  
starburst nucleus. }
\label{hi_tau2} 
\end{figure}

As noted above, it is well known that the m/cm-wave recombination lines ($n > 100$) of HI have substantial negative optical 
depths and hence, maser amplification of the line emission. In such instances, the recombination line intensity will not 
accurately reflect the ionized gas emission measure and the associated Lyman continuum emission rates of the stellar population. 
For the submm HI and HeII lines we can analyze the possibility of maser amplification using the optical depth information 
of \cite{sto95}. They provide an optical depth parameter $\Omega_{n,n'}$ which is related to the line center optical depth $\tau_{n,n'}$  by 

\begin{eqnarray}
\tau_{n,n'} =  n_e n_{ion} \Omega_{n,n'} L ,
\end{eqnarray}

\noindent where L is the line-of-sight path length. $\Omega_{n,n'}$ is inversely proportional to the line width in Hz, $\Delta_{n,n'}$
and in their output they used a thermal doppler width, implying a velocity full width at half maximum intensity 
\begin{eqnarray}
\Delta v_{FWHM} &=& \left( {8~\ln 2 ~k T_e \over{m_{ion} }}\right)^{1/2} 
\end{eqnarray}
\noindent or 21.7 km sec$^{-1}$ for HI at 10$^4$ K. In most situations relevant to the discussion here, the line widths will exceed the thermal width due to large scale bulk motions within 
the host galaxies. We have therefore rescaled the optical depths to $\Delta v_{FWHM} = 100$ km sec$^{-1}$. We have also scaled the optical 
depth to a {\it specific} optical depth $\tau$ per unit $n^e n_{ion} L_{pc}$ where $L_{pc}$ is the path length in parsecs and the volume densities in cm$^{-3}$.

Figures \ref{hi_tau1} and  \ref{hi_tau2} show the {\it specific} optical depths for the HI and HeII lines as a function of $T_e$ and $n_e$. The actual optical depths for a 
particular source may be obtained by scaling these curves by the factor $n_e n_p L_{pc} / \Delta v_{100}$.  In these plots, the dashed lines are 
for transitions with a population inversion and hence negative optical depth. 

The submm transitions of HI and HeII are principle quantum number $n \sim 20$ to 32 and 32 to 50, respectively. 
For both HI and HeII these particular transitions have positive specific optical depths and hence no maser amplification at virtually all densities and temperatures shown in Figures \ref{hi_tau1} and  \ref{hi_tau2}.
The exceptions to this are that at 
very high densities, $n_e > 10^6$ cm$^{-3}$, there can be population inversions (see Figure \ref{hi_tau2}). However, even at these high densities, 
significant amplification would occur only if the scale factor is sufficiently large. 

An extreme upper limit for the HII in a ULIRG starburst 
nucleus might be $n_e n_p L_{pc} \sim 10^4\times10^4\times10 = 10^9$ cm$^{-6}$ pc and $\sim 10^8$ cm$^{-6}$ pc for HeII. Applying the first scale factor to the 
curves shown in Figure \ref{hi_tau2} yields upper limits to the negative optical depth $| \tau | < 0.1$, implying insignificant amplification even for these extreme conditions. 

In summary, the observed emission line fluxes for the submm recombination lines will 
provide a linear probe of the HII and HeIII EMs; they will not be affected by non-linear radiative transfer effects, either maser amplification or optically thick saturation of the emission.  

As an aside, it is interesting to note that the behavior of the HI opacities shown in Figure \ref{hi_tau1}-left is reflected in the HeII 
opacities (Figure \ref{hi_tau1}-left ) but translated to higher n$\alpha$ transitions. This is of course expected since HeII is hydrogenic and the energy levels 
are scaled by a factor 4, implying higher principal quantum number in HeII to obtain similar line frequencies and A coefficients. 

\subsection{Excitation by Continuum Radiation in Lines ?}\label{cont}

Lastly, we consider the possibility that absorption and stimulated emission could alter the bound level populations away from 
those of a radiative cascade following recombination. \cite{wad83} analyzed this effect on the cm-wave recombination lines in 
powerful, radio-bright QSOs. For the submm lines considered here, the radiative excitation would be provided by the infrared 
continuum. Significant coupling of the level populations to the local radiation field at the line frequencies occurs when the net radiative excitation rate (i.e. absorption minus stimulated 
emission) is comparable with the spontaneous decay rate.  It is easily shown that this happens when the local energy density of the radiation field exceeds that of a black body with temperature greater than $T_x$, where 
$T_x$ is the excitation temperature characterizing the cascade level populations. (This is the radiative equivalent of the critical density often used to characterize the collisional coupling of 
levels to the gas kinetic temperature.) Neglecting departures from thermal equilibrium and letting $T_x \sim 10^4$ K, the effective radiation 
temperature $T_R$ must be therefore be $> 10^4$ K at the submm line frequencies. 

This scenario is probably only of conceivable relevance for an AGN and not for a starburst. 
For example, suppose the AGN luminosity is 
$\sim 10^{12}$ \lsun, then the effective black body radius for $10^4$ K is 0.007 pc. Inside this radius the radiation energy density will exceed that of a $10^4$ K black body, but at 
larger radius the induced radiative transitions become much less important. For the ionization case of an AGN as discussed below (\S \ref{ioniz}), the radii of the HeIII and HII regions are 16 and 27 pc respectively, as shown in Figure \ref{ioniz_agn}.  For this very simplified example, we do not therefore expect radiative excitation in the bound-bound transitions to be significant in the bulk of the ionized gas. For other instances, one can easily perform a similar analysis as a check.

\section{Ionization Structure of Starbursts and AGN Sources}\label{ioniz}

To evaluate the expected line luminosities for the HI and HeII lines, we now calculate the ionized gas
emission measures for the typical EUV spectra associated with starbursts and AGN. With the derived EMs for H$^+$ and He$^{++}$ as scale factors for HI and HeII 
emissivities per unit EM (\S \ref{he_emiss} and \ref{h_emiss}), one can then calculate the line luminosities.

\begin{figure}
\epsscale{1.0}  
\plotone{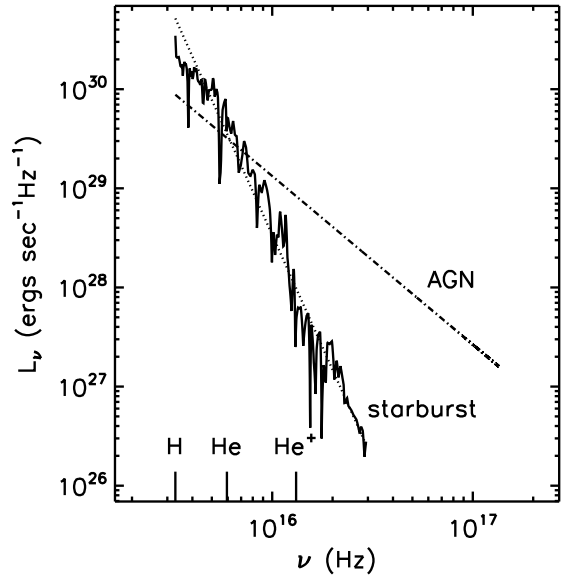}
\caption{The input ionizing continua are shown for starburst and AGN sources. The starburst EUV down to $\lambda = $90\AA~ was computed from Starburst99 with a Kroupa 
IMF, solar metallicity and a constant SFR. The AGN EUV continuum was taken as a power-law with specific luminosity index $\nu^{-1.7}$. Both EUV spectra were normalized so that 
the integrated EUV luminosity (at $\lambda < 912$ \AA) was 10$^{12}$ \lsun. 
The ionization thresholds for H, He and He$^{+}$ are shown as vertical lines on the bottom axis. The dotted line along the SB spectrum is a power-law fit to 
the Starburst99 spectrum with L$_{\nu} \propto \nu^{-4.5}$ used for the analytic treatment in \S \ref{anal}.}
\label{ioniz_uv} 
\end{figure} 

For the starburst (SB) spectrum, we adopt a Kroupa IMF (0.1 to 100 \msun) and use the Starburst99 spectral synthesis program \citep{lei99} to calculate the EUV 
spectrum at solar metallicity for a continuous SFR. After 10 Myr, the EUV at $\lambda < 912$ \AA~ is constant since the early type star population 
has reached a steady state with equal numbers of new massive stars being created to replace those evolving off the main sequence. This EUV continuum can then be taken to represent  
a steady state SFR -- applicable to starbursts lasting more the 10$^7$ yrs. For the AGN EUV-X-ray continuum, we adopt a simple power-law $L_{\nu} \propto \nu^{-1.7}$ \citep[e.g.][]{ost06}.
We scale both the SB and AGN $L_{\nu}$ to have integrated EUV luminosity $= 10^{12}$ \lsun at $\lambda < 912$\AA. For the starburst spectrum, this EUV luminosity
corresponds to a steady-state SFR $= 874 $ \msun yr$^{-1}$ for a Kroupa IMF. These two ionization spectra are shown in Figure \ref{ioniz_uv}. 
The figure clearly demonstrates the significant difference between the SB and AGN EUV spectra, with the former having almost no photons 
capable of ionizing He$^+$ to He$^{++}$, compared to the number of HI ionizing photons. 

Using these EUV continua, we have computed the ionization structure for a cloud with H density $n = 10^4$ cm$^{-3}$, assuming all EUV photons 
are used for ionization, i.e. the plasma is ionization bounded and no EUV is absorbed by dust within the ionized gas. The He/H abundance ratio was 0.1.
The EUV continuum was assumed to originate in a central point source and the specific luminosity of the ionizing photons at each radius was attenuated 
by the optical depth at each frequency due to H, He and He$^+$ along the line of sight to the central source. The secondary ionizing photons produced 
by recombinations with sufficient energy to ionize H or He were treated in the 'on the spot' approximation, i.e. assumed to be absorbed at the radius they 
were produced. Lastly, we simplified the analysis of these secondary photons by assuming a fraction 0.96 and 0.66 of the He$^{+}$ recombinations yielded 
a photon which ionizes HI at electron densities below and above 4000 cm$^{-3}$ \citep[see][]{ost06}, respectively.

\begin{figure}
\epsscale{1.0}  
\plotone{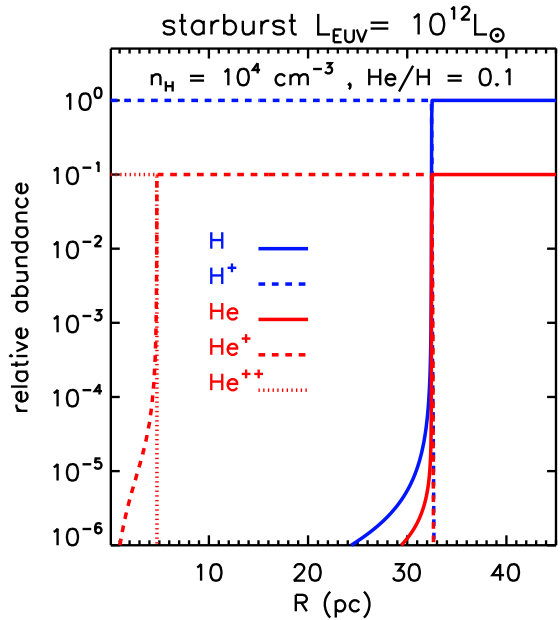}
\caption{For a starburst EUV spectrum with L$_{EUV} = 10^{12}$ \lsun and Hydrogen density n = $10^4$ cm$^{-3}$, the radii of the H$^+$, He$^+$ and He$^{++}$ regions are
shown. The He/H abundance ratio was 0.1.}
\label{ioniz_sb} 
\end{figure}

\begin{figure}
\epsscale{1.0}  
\plotone{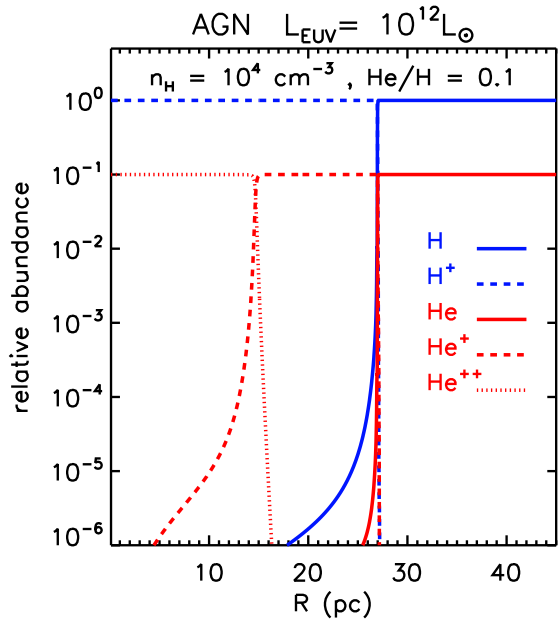}
\caption{For an AGN power-law EUV-X-ray spectrum with L$_{EUV} = 10^{12}$ \lsun and Hydrogen density n = $10^4$ cm$^{-3}$, the radii of the HII, HeII and HeIII regions are
shown. The He/H abundance ratio was 0.1.}
\label{ioniz_agn} 
\end{figure}

 Figures \ref{ioniz_sb} and \ref{ioniz_agn} show the 
relative sizes of the HII, HeII and HeIII regions for the SB and AGN. These figures clearly show the marked contrast in size (and hence EM) of the He$^{++}$
regions in the two instances.  Much less contrast is seen in the He$^{+}$ emission measures between the two models. 

From ionization equilibrium calculations for the SB and AGN EUV spectra, we derive the EM of the ionized regions in HII, HeII and HeIII (Table \ref{tab1}). 
For both ionizing sources the spectra were normalized to have L$_{EUV} = 10^{12}$ \lsun. 
From the EMs shown in Table \ref{tab1}, we draw two important conclusions: 1) despite the very different spectral shapes, the bulk of the 
ionizing continuum is absorbed in the HII region and the EM$_{H^+}$ provides a reasonably accurate estimate of the total EUV luminosity, differing 
less than a factor 2 between the two cases, and 2) the EM ratio, EM$_{He^{++}}$/EM$_{H^+}$, is 50 times greater for the AGN than for the SB, indicating 
that this ratio provides an excellent diagnostic of AGN versus SB ionizing sources.

\begin{deluxetable*}{cccccccc}

\tabletypesize{}
\tablecaption{\bf{Emission Measures of Ionized Regions} \label{tab1}  }
\tablewidth{0pt}
\tablehead{\colhead{EUV} & \colhead{L$_{EUV}$}& \colhead{SFR/$\dot{M}_{acc}$}\tablenotemark{a} & \colhead{Q$_{Lyc}$} & \colhead{n$_e$n$_{p}$vol} & \colhead{n$_e$n$_{He^+}$vol} & \colhead{n$_e$n$_{He{}^{++}}$vol } &  \colhead{L$_{H26\alpha}$} \\
\colhead{} & \colhead{} & \colhead{} & \colhead{} & \colhead{} & \colhead{} & \colhead{}  & \colhead{/L$_{HeII42\alpha}$} \\
\colhead{} &  \colhead{\lsun} & \colhead{\msun/yr} & \colhead{sec$^{-1}$} & \colhead{cm$^{-3}$} & \colhead{cm$^{-3}$}  & \colhead{}  }
\startdata
Starburst & $10^{12}$ &674 & $1.20\times10^{56}$& $4.61\times10^{68}$ & $ 4.59\times10^{67}$ &$1.59\times10^{65}$ & 0.0017\tablenotemark{b} \\
AGN & $10^{12}$ & 0.65 & $6.97\times10^{55}$ &  $2.68\times10^{68}$ & $ 2.21\times10^{67}$ &$4.64\times10^{66}$ &  0.084\tablenotemark{b} \\
\enddata
\tablecomments{L$_{\nu}$ for the SB and AGN were normalized to both have 10$^{12}$ \lsun in the Lyman continuum at $\lambda < 912$ \AA (Column 2). Based on comparing Figures \ref{hi_emiss_submm} and \ref{he_emiss_submm}, 
the HeII-42$\alpha$ line has 4.82 times larger flux per unit EM than the H-26$\alpha$ line; this factor is used to estimate the line luminosity ratio given in column 8. Line ratio is for T = 10$^4$K. The emission measures (EM) 
are given in cm$^{-3}$.}
\smallskip
\tablenotetext{a}{SFR or AGN accretion rate required to give this $L_{EUV}$. The SFR assumes a Kroupa IMF; it would be  factor 1.6 
 higher for a Salpeter IMF. The AGN accretion rate assumes 10\% of the mass accretion rate is converted to EUV luminosity.}
\tablenotetext{b}{H-26$\alpha$ and HeII-42$\alpha$ are at 353.623 and  342.894 GHz respectively, separated by 10.7 GHz
and therefore observable in a single tuning with ALMA Band 7. }
\end{deluxetable*}

\subsection{Ionization Structure -- Analytics}\label{anal}

In the previous section, we made use of a full ionization equilibrium model using Starburst99 for the starburst EUV spectrum
and a power law approximation for the AGN EUV. In this numerical treatment, we track the competition of all three species (HI, HeI and HeII) for 
ionizing photons at each wavelength. However, an analytic 
treatment, which turns out to reproduce quite well the full numerical approach, can be developed using 
a few simplifying assumptions regarding the EUV spectra and the competition between the 3 species for the 
ionizing photons in the energy regimes above 24.6 ev. 

At photon energies between 24.6 to 54.4 eV, both HI and HeI can be ionized, and at energies above 54.4 eV all three species (HI, HeI and HeII) 
can be ionized. However, for all three species, the ionization cross sections are highest at the thresholds and drop 
as $\nu^{-3}$ above their respective thresholds. At 24.6 eV (the HeI ionization threshold), the HeI ionization cross section is 8 times larger than 
that of HI at the same energy. Thus, provided HI is mostly ionized, the photons above this energy are largely used to ionize HeI. These two factors 
(the higher cross section and the fact that HI is mostly ionized) more than make up for the fact that the He/H abundance ratio 
is 0.1 (see Figures \ref{ioniz_sb} and \ref{ioniz_agn}). A fraction of the HeI 
recombination photons can ionize HI so effectively that each of the photons above 24.6 eV will ionize both HeI and HI. This fraction varies between 
0.96 (for $n_e < 4000$ cm$^{-3}$) and 0.66 (for $n_e > 4000$ cm$^{-3}$) \citep[see][]{ost06}. Thus, one can approximate the number of photons 
available to ionize HI as all those above 13.6 eV. Above 54.4 eV, the photons are predominantly used for HeII ionization since the abundance of 
HI will be very low in the HeII region. (This last assumption presumes that the ionizing continuum is hard enough that the gas is easily ionized to HeII.) 

In comparing the ionization associated with a SB versus an AGN, we assume that the ionized regions are ionization bounded and that 
dust within the ionized regions does not significantly deplete the EUV. The latter could be a significant issue for very young ionized regions 
but is perhaps less likely for SB and AGN ionized regions where the timescales are $> 10^7$ yrs and the dust within the ionized 
gas is likely to have been destroyed. Under these assumptions, the standard Str\"{o}mgren relation implies that the total volume integrated 
emission measure of each species ionized region will be determined by the total production rate of fresh ionizing photons. For comparing 
the SB and AGN cases, we normalize both EUV spectra to have the same total integrated EUV luminosity, 

\begin{equation}
  L=\int \limits^\infty_{\nu_{th}} L_{\nu} d\nu ,
   \end{equation}
   
\noindent where $\nu_{th}$ is the ionization threshold frequency. The production rate of ionizing photons is then given by Q, with 
  
  \begin{equation}
Q = \int \limits^\infty_{\nu_{th}} L /h\nu ~d\nu .
  \end{equation}

We consider the three regimes in the EUV 
 \begin{enumerate}
\item h$\nu \geq13.6$ eV  and h$\nu_{th}$ = h$\nu_0$ = 13.6 eV 
\item h$\nu \geq 24.6$ eV and h$\nu_{th}$ = h$\nu_1$ = 24.6 eV 
\item h$\nu \geq 54.4$ eV and h$\nu_{th}$ = h$\nu_2$ = 54.4 eV  
\end{enumerate}

\noindent corresponding to H ionization, He to He$^{+}$ ionization and He$^+$ to He$^{++}$ ionization. 

For the EUV spectra we make the assumption that {\it both} SB and AGN EUV spectra can be represented by 
power-laws with 
$L_{\nu} = C_{SB} \nu^{-\alpha_{SB}} $ and $C_{AGN} \nu^{-\alpha_{AGN}}$. For the AGN, this is a commonly used assumption with $\alpha_{AGN} = 1.7$. For the SB this assumption may appear surprising but 
Figure \ref{ioniz_uv} clearly shows that the EUV spectrum obtained from the spectral synthesis of a continuous SB can be fit by a power-law with $\alpha_{SB} = 4.5$.
For these simple power-laws, the luminosity normalization yields the relation

  \begin{equation}
{C_{AGN}\over{C_{SB}}}  =   {( 1 - \alpha_{AGN}) \over{ ( 1 - \alpha_{SB}) }} \nu_0^{-\alpha_{SB}+\alpha_{AGN}} 
  \end{equation}

  \noindent and for $ \alpha_{AGN} = 1.7$ and $\alpha_{SB} = 4.5$, this reduces to $C_{AGN} / C_{SB} = 5 \nu_0^{2.8}$.
  
The Str\"{o}mgen ionization equilibrium for a power-law ionizing spectrum then yields 
  \begin{equation}
{Q_{He^{++}}\over{Q_{H^+}}}  =   \left( {\nu_2 \over{\nu_0}} \right) ^{-\alpha} = 4^{-\alpha}
  \end{equation}\label{ratio}

\noindent since $\nu_2 = 4 \nu_0$. For the AGN with $\alpha_{AGN} = 1.7$,  $Q_{He^{++}} / Q_{H^{+}} = 0.095$ and for the SB, 
$\alpha_{SB} = 4.5$,  $Q_{He^{++}} / Q_{H^{+}} = 1.95\times10^{-3}$.

For Case B recombination, the Qs are related to the emission measures of their respective Str\"{o}mgren spheres by 
the recombination coefficients to states above the ground state and the electron density 
\begin{eqnarray}
 Q_{H^{+}} &=& n_e n_p ~vol~ \alpha_b(H) \nonumber \\
 &\simeq& 2.60\times10^{-13} n^2 ~vol~ (T/10^4)^{1/2}
\end{eqnarray}

\noindent and
\begin{eqnarray}
Q_{He^{++}} &=&   n_e n_{He^{++}} ~vol~ \alpha_b(He^+) \nonumber \\
 &\simeq& 1.85\times10^{-12} n n_{He^{++}} ~vol~ (T/10^4)^{1/2} ,
\end{eqnarray} \label{hep}

\noindent  where n is the number density of H nuclei, vol is the volume of the ionized region and we set $n_e$ = 1.1 and 1.2$\times n_H$ for the H$^+$ and He$^{++}$ regions respectively. 

For T $= 10^4$K and [He/H] = 0.1, the emission measures are 

\begin{eqnarray}
EM_{H^+} &=& Q_{H^{+}} / 2.60\times10^{-13}  \nonumber \\
 &=& 3.85\times10^{12} Q_{H^{+}} 
\end{eqnarray}

\noindent and
\begin{eqnarray}
EM_{He^{++}} &=& Q_{He^{++}} / 1.85\times10^{-12} \nonumber \\
&=& 5.41\times10^{11} Q_{He^{++}} .
\end{eqnarray}
  
\noindent The emission measure ratio is therefore 
\begin{equation}
EM_{He^{++}}/EM_{H^{+}} = 0.141 Q_{He^{++}}/Q_{H^{+}} .
\end{equation}

Thus, $EM_{He^{++}}/EM_{H^{+}} = 1.34\times10^{-2}$ and $2.75\times10^{-4}$ for the AGN and SB EUVs, respectively.
From the results of the numerical calculation given in Table \ref{tab1}, the ratios were $1.73\times10^{-2}$ and $3.45\times10^{-4}$, respectively. 
We therefore conclude that the simple analytic approach provides excellent agreement with the results quoted above for full numerical ionization equilibrium calculation obtained using the detailed SB99 spectrum for the starburst. 

Lastly, we note that the change in the AGN / SB ratio of EMs ($He^{++}/H^+$) is easily shown from Eq. \ref{ratio} to 
be 
  \begin{equation}
 { EM(He^{++}/H^+)_{AGN} \over EM(He^{++}/H^+)_{SB}} = 4^{-\alpha_{AGN} + \alpha_{SB}} = 48.5 
  \end{equation}
\noindent   as compared with 49.4 from the numerical analysis above. Contrasting this large change in the He$^{++}$ between the 
SB and AGN EUVs, Table \ref{tab1} shows only $\sim10$\% change in the ratio of He$^+$ relative to H$^+$ between the two 
cases, implying that the HeI recombination lines can not be used to discriminate AGN and SB EUVs.


  \begin{deluxetable*}{ccccrrr}{ht}
\tablecaption{\bf{HI and HeII Paired Submm Lines} \label{tab2}  }
\tablewidth{0pt}
\tablehead{\colhead{HI } & \colhead{$\nu$ } & \colhead{HeII n$\alpha$} & \colhead{$\nu$ (GHz)} &  \colhead{$\Delta \nu$ (GHz)} & \colhead{$\epsilon _{HI}$} & \colhead{$\epsilon _{HeII}$}  \\
\colhead{n$\alpha$} & \colhead{GHz} & \colhead{n$\alpha$} & \colhead{GHz}  & \colhead{GHz}  & \colhead{erg sec$^{-1}$ cm$^3$} & \colhead{erg sec$^{-1}$ cm$^3$} }
\startdata
20  &  764.230 &  32 & 766.940 &  -2.710 & 1.21$\times10^{-30}$ & 4.45$\times10^{-30}$ \\
21 &   662.404 &  34 & 641.108 &  21.296 & 9.05$\times10^{-31}$ & 3.09$\times10^{-30}$\\
22 &   577.896 &  35 & 588.428 & -10.531  & 6.85$\times10^{-31}$ & 2.59$\times10^{-30}$\\
23 &   507.175 &  37 & 499.191 &   7.985 & 5.25$\times10^{-31}$ & 1.85$\times10^{-30}$ \\
24 &   447.540 &  38 & 461.286 & -13.746  & 4.06$\times10^{-31}$ & 1.57$\times10^{-30}$\\
25 &   396.901 &  40 & 396.254 &   0.647  & 3.17$\times10^{-31}$ & 1.15$\times10^{-30}$\\
26 &   353.623 &  42 & 342.894 &  10.729 & 2.50$\times10^{-31}$ & 8.47$\times10^{-31}$ \\
27 &   316.415 &  43 & 319.781 &  -3.366  & 1.99$\times10^{-31}$ & 7.32$\times10^{-31}$\\
28 &   284.251 &  45 & 279.432 &   4.818  & 1.60$\times10^{-31}$ & 5.50$\times10^{-31}$\\
29 &   256.302 &  46 & 261.787 &  -5.485  & 1.29$\times10^{-31}$ & 4.79$\times10^{-31}$\\
30 &   231.901 &  48 & 230.713 &   1.187  & 1.05$\times10^{-31}$ & 3.67$\times10^{-31}$\\
31 &   210.502 &  50 & 204.370 &   6.132  & 8.67$\times10^{-32}$ & 2.83$\times10^{-31}$ \\
32 &   191.657 &  51 & 192.693 &  -1.036  & 7.18$\times10^{-32}$ & 2.48$\times10^{-31}$\\
\enddata
\tablecomments{In ALMA Band 7 (275 to 365 GHz), the IF frequency is 4 GHz and the correlator has a nominal coverage of 
4 $\times$ 2 GHz or 8 GHz in each sideband. Therefore a single tuning can cover $\sim$16 GHz of bandwidth.} 
\end{deluxetable*}

As an aside, we note that we were surprised to find that the SB99 EUV spectrum shown in Figure \ref{ioniz_uv} could be fit by a power-law. Upon investigating this 
  further, we found that there is enormous variation in the model EUV spectra depending on which stellar 
  atmosphere model was employed, and due to the very uncertain contributions of Wolf-Rayet stars. Given these large uncertainties
  in the predicted SB EUV spectra, the specific power law index adopted above should only be taken as illustrative. Instead, 
  it would be more appropriate to take the power-law as a 'parameterization' which allows simple exploration of the EUV 
  spectral properties and HI and HeII emission line ratios. In fact, measurements of the $EM(He^{++}/H^+)$ ratio might be used to 
  constrain the very uncertain EUV spectra of SB regions and OB star clusters. An alternative parameterization might be to 
  model the SB EUV as a blackbody. For $T_{BB} = 45,000$K, 
  $Q_{He^{++}}/Q_{H^+} \simeq 2.8\times10^{-4}$, implying a similar ratio for the emission measures.  This is effectively a factor 10
  lower than the ratio obtained for SB99 EUV and the $\alpha_{SB}=4.5$ power-law used above.

  \section{Paired HI and HeII Recombination Lines}

In Table \ref{tab2} we provide a list of the submm HI recombination lines together with their closest 
frequency-matched HeII lines. The ALMA IF frequency is 4 GHz and each correlator has a maximum bandwidth of 
1.8 GHz, thus in a single tuning the spectra can cover up to 16 GHz. One prime pairing for simultaneous coverage of HI and HeII
occurs at 350 GHz where HI-26$\alpha$ and HeII-42$\alpha$ can be observed within a good atmospheric window (ALMA Band 7). 
In Table \ref{tab1}, the last column gives the expected line ratio, HeII-42$\alpha$/HI-26$\alpha$ derived from the EM given in Table \ref{tab1}. 
The emissivities are shown in Figures \ref{hi_emiss_submm} and \ref{he_emiss_submm}. The line ratio varies by a factor 50 between the two cases, 
clearly demonstrating the efficacy of the HeII/HI submm line ratios to discriminate the nature of the ionizing sources. By contrast, 
the ratio EM$_{H^+}$/EM$_{He^+}$ is different only by a factor 10\% between the SB and AGN cases, indicating that the HeI/HI recombination 
line ratios are not a good SB versus AGN discriminant. 
   
 \section{Star Formation Rates and AGN Luminosity}
 
Derivation of SFRs and AGN accretion rates from the HI and HeII recombination lines are potentially quite straightforward provided
the form of each EUV spectrum can be parameterized. For the preceding analysis we normalized the EUV luminosity to $10^{12}$ \lsun for both the SB and AGN. For a continuous SB (extending 
 over $> 10^7$ yrs, the EUV luminosity will be constant. This EUV luminosity ($10^{12}$ \lsun) translates to the steady state SFR $= 874 $ \msun yr$^{-1}$ for a Kroupa IMF. The implied SFR is a factor 1.6 
 higher for a Salpeter IMF. (The total stellar luminosity integrated over all wavelengths would be $5.5\times10^{12}$ \lsun at $10^7$ yrs.) For an AGN 
 with L$_{EUV} = 10^{12}$ \lsun, this EUV luminosity corresponds to an accretion rate of 0.65 \msun yr$^{-1}$ assuming 10\% conversion of accreted mass to EUV photon energy. 
 [Note that the above luminosities refer to that in the EUV, not the total bolometric luminosities.]
 
 For galaxies with these luminosities, the submm recombination lines of both HI and HeII are detectable with 
 ALMA out to distance $\sim 100$ Mpc in a few hours integration. As an example, consider the H-$26\alpha$ line in a ULIRG like Arp 220 (or NGC 6240) 
 at a distance $\sim100$ Mpc with an HII emission measure EM$_{H^+}$ (see Table \ref{tab1}). 
For a specific emissivity $\epsilon$, emission measure EM and source distance D (all in cgs units), the velocity-integrated line flux in observer units Jy km sec$^{-1}$ is given by
 
 \begin{eqnarray}
 S \Delta V = {\epsilon  ~EM_{H^+} \over{ 4\pi D^{2}}} {c \over{\nu_{obs}}} 10^{18} {\rm ~Jy~km~sec}^{-1}  \nonumber 
 \end{eqnarray}
 
\noindent  Using the volume emissivity of $\epsilon = 2.5\times10^{-31}$ ergs cm$^{-3}$ sec$^{-1}$ / $n_e n_p vol$ from 
 Figure \ref{hi_emiss}, one finds the velocity-integrated line flux for H-$26\alpha$:
 
 \begin{eqnarray}
 S_{HI-26\alpha} \Delta V &=& 7.17 {EM_{H^+} \over{4\times10^{68}}} D_{100Mpc}^{-2} \nu_{350GHz}^{-1} \nonumber  \\
 &&{\rm ~Jy~km~sec}^{-1}.
 \end{eqnarray}

\noindent The frequency-paired HeII$-42\alpha$ line will have an integrated flux $\sim8$\% of HI$-26\alpha$ in the case of AGN. Both 
lines should be simultaneously detectable in a few hours with ALMA. For reasonable densities, the emission in these lines 
will be directly proportional to the EM of the gas. Even at $n_e = 10^6$ cm$^{-3}$, the emission rate in the HI$-26\alpha$ and 
 HeII$-42\alpha$ lines are altered by only 1 and 2\%, respectively. If the source is known to be a 'continuous' starburst, one may substitute SFR/(674 \msun yr$^{-1}$) for 
 EM$_{H^+}/(4\times10^{68})$ in the equation above,
 
 \begin{eqnarray}
 S_{HI-26\alpha} \Delta V &=& 1.06 {SFR \over{100 \msun yr^{-1}}} D_{100Mpc}^{-2} \nu_{350GHz}^{-1} \nonumber  \\
 &&{\rm ~Jy~km~sec}^{-1}.
 \end{eqnarray}

  We have recently detected the HI$-26\alpha$ in Arp 220 in ALMA Cycle0 observations 
 with a line flux indicating a $SFR \simeq 100$ \msun yr$^{-1}$ (Scoville \etal 2013 -- in preparation). \cite{yun04} also report detection of HI$-41\alpha$
 at 90 GHz yielding a similar SFR. 
 
 As noted earlier, the free-free (Bremsstrahlung) continuum emission can also be used to probe the ionized gas EM. For completeness, the free-free flux 
 density in the submm regime is given by 
  \begin{eqnarray}
 S_{ff}  = 75.0 {EM_{H^+} \over{4\times10^{68}}} ~\nu_{300 Ghz}^{-0.17} ~T_{10^4 K}^{-0.5} ~D_{100Mpc}^{-2} \rm ~mJy,
 \end{eqnarray}
\noindent where we have included a factor 1.1 to account the the He$^+$ free-free emission assuming [He/H] = 0.1 and $\nu^{-0.17}$ is the frequency dependence of the Gaunt factor at submm wavelengths.  In most instances the 
thermal dust emission will dominate the free-free so the latter is not generally a useful tracer of the ionized gas EM. 
 
 \subsection{Dust Extinction}
 
We have stressed that a major advantage of the submm recombination lines is that they are at sufficiently long wavelengths that 
dust extinction should be negligible, since for standard dust properties the extinction should be A($\lambda)\sim10^{-4}( \lambda _{\mu m} / 300\mu m)^{-1.8} A_V$
at submm and longer wavelengths \citep[e.g.][]{bat11,pla11a,pla11b}. Thus for A$_V < 1000$ mag, extinction at $\lambda > 300 \mu m$ should be minimal. However, there are 
a few extreme cases such as Arp 220 and young protostellar objects which may have somewhat higher dust columns. In these cases, the recombination 
lines provide a unique probe of the dust extinction through measurements of HI lines at different submm wavelengths. Their intrinsic flux ratios can be determined from 
Figure \ref{hi_emiss_submm}; the extinction is then obtained by comparison of the intrinsic and observed line ratios. 
In sufficiently bright recombination line sources with high extinction, such observations could potentially be used to determine the 
frequency dependence of the dust extinction in the submm -- this has been a major uncertainty in the analysis of submm continuum observations. 

\section{Conclusions}

We have evaluated the expected submm wavelength line emission of HI, HeI and HeII as probes of dust embedded star formation 
and AGN luminosity. We find that the low-n $\alpha$ transitions should provide a linear probe of the emission measures 
of the different ionized regions. Although their energy levels will have population inversions, the negative optical
depths will be $<< 0.1$ for the maximum gas columns expected and hence there is no significant maser amplification. 

The submm
HI and HeII lines have major advantages over other probes of SF and AGN activity: 1) the dust extinctions should be minimal; 2) the 
emitting levels (n $< 30$ for HI and $< 50$ for HeII) have high critical densities \citep[n$_{crit} > 10^4$ cm$^{-3}$,][]{sej69} and hence will not be collisionally suppressed; and 3) they arise from the most abundant 
species and therefore do not have metallicity dependences. The emission line luminosities of the HI (and HeI) submm recombination lines are therefore 
a direct and linear probe of the EUV luminosity and hence SFR if the source is a starburst.

The emission ratios of HI to HeII can be a sensitive probe of the 
hardness of the EUV ionizing radiation field, providing a clear discriminant between AGN and SBs. 

The observed ratios of the submm HI recombination lines may also be used to determine the extinction 
in highly extincted luminous sources and to constrain the shape of the submm extinction curve. 

Lastly, we find that 
these lines should be readily detectable for imaging with ALMA in luminous galaxies out to 100 Mpc and 
less luminous sources at lower redshift.  We note that the far infrared fine structure lines observed with Herschel often show line deficits in the ULIRGs relative to the IR luminosity, 
possibly indicating either dust absorption of the EUV or collisional suppression of the emission rates at high density \citep{gra11}; the latter will not be a problem since the HI and HeII lines are permitted transitions with high spontaneous decay rates. 

\acknowledgments
We thank Chris Hirata for discussions during this work, Zara Scoville for proof reading the manuscript and Jin Koda and Min Yun for comments.
We thank the Aspen Center for Physics and the NSF Grant \#1066293 for hospitality during the writing of this paper. We also thank the referee 
for suggesting we include a discussion of radiative line excitation (\S \ref{cont}). 


\appendix

\let\a=\alpha \let\b=\beta \let\g=\gamma \let\d=\delta \let\e=\epsilon
\let\z=\zeta \let\h=\eta \let\th=\theta \let\i=\iota \let\k=\kappa
\let\l=\lambda \let\m=\mu \let\n=\nu \let\x=\xi \let\p=\pi \let\r=\rho
\let\sig=\sigma \let\t=\tau \let\u=\upsilon \let\f=\phi \let\c=\chi \let\y=\psi
\let\vp=\varphi \let\vep=\varepsilon
\let\w=\omega	   \let\G=\Gamma \let\D=\Delta \let\Th=\Theta \let\L=\Lambda
\let\X=\Xi \let\P=\Pi  \let\U=\Upsilon \let\Y=\Psi
\let\C=\Chi \let\W=\Omega

\def \z {\xi}
\def \x {\chi}

\def\td{\tilde} \def\wtd{\widetilde}
\def \const {{\rm const}}
\def \del {\partial}

\let\la=\label \let\ci=\cite
\def\no{\nonumber \displaystyle } \def \foot {\footnote}
\def\ni{\noindent}
\def \bi{\bibitem}
\def \size {\displaystyle}

\def\ba{\begin{array}} \def\ea{\end{array}}
\def \be {\begin{eqnarray}} \def\ee{\end{eqnarray}}
\def \bei {\begin{itemize}} \def \eei {\end{itemize}}

\def \half {\frac{1}{2}}

\def \He {{\rm He}}
\def \H {{\rm H}}

\def \LL {{\cal L}}
\def \QQ {{\cal Q}}

\def\agn {{\rm AGN}}
\def\bb {{\rm BB}}
\def \lum {luminosity}
\def \ev {{\, \rm eV}}
\def \cm {{\rm cm}}
\def \pc {{\rm pc}}
\def \Ry {{\rm Ry}}
\def \io {{\rm io}}
\def \ell {{\frak l}}

\def\br {{\bf r}}
\def \la {\langle}
\def \ra {\rangle}

\section{Scaling Relations for Hydrogenic Ions}\label{appen_scale}

The aim of this appendix is to lay the basis of partial analytical explanation of HeII to HI emissivities scaling relation discussed previously. To this end, in this following sections of this Appendix, we analyze radiative processes involving Hydrogen-like atom consisting of Z charged nucleus and single electron orbiting it 
and put them in use in Appendix B. In this notations HI corresponds to $Z=1$ and HeII to $Z=2.$

Below we refer to \cite{lan77} and \cite{ber82}, as examples of standard course in quantum mechanics and QED. The choice is dictated by our personal preferences. The reader is free to refer to any standard textbook in quantum mechanics and QED or original papers, references to which can be found in \cite{lan77} and \cite{ber82}.   ~
\medskip 

\subsection{Radiation}\label{acoef}

Scaling of the Einstein A-coefficients for 
spontaneous radiative decay can be explicitly derived in the dipole approximation.
The probability of dipole transitions between two states of the hydrogenic ion is given by
\begin{equation}\label{1.1}
	A_{n_1\to n_2} =  \frac{4 (\w_{n_1\to n_2})^3}{3 \hbar c^3} d_{n_1\to n_2}^2,
\end{equation}
where $w_{n_1\to n_2}  = Z^2 \frac{m e^4}{2\hbar^3}\left(\frac{1}{n_2^2}-\frac{1}{n_1^2}\right)=  Z^2 w^H_{n_1\to n_2}$ is the frequency of radiated photon and
$d^2_{n_1\to n_2} $ is the average over $l'$s and $m'$s of the transition dipole moment 
$d_{n_1l_1m_1\to n_2l_2m_2}.$
Here
\begin{eqnarray}
&& \size
	d^2_{n_1\to n_2} = \frac{1}{n^2} \sum\limits_{l_1, m_1, l_2, m_2} 
	d^2_{n_1l_1m_1\to n_2l_2m_2}\quad {\rm and} \quad d^2_{n_1l_1m_1\to n_2l_2m_2} =\la d_x\ra^2+\la d_y\ra^2+\la d_z\ra^2
\end{eqnarray}
with $\la d_i \ra = -\la \psi_{n_2 l_2 m_2} |  e r_j  | \psi_{n_1l_1m_1}\ra, \ j=x,y,z $ and
 $r_j$ is the component of electron radius vector in atom (we neglect the motion of nucleus).
The wave functions $\psi_{in}=\psi_{n_1 l_1 m_1}$ and $\psi_{f}=\psi_{n_2 l_2 m_2}$ are the initial and final wave functions of the electron on $n_1l_1m_1$ and $n_2l_2m_2$ levels of hydrogenic ion.

One can show (see any standard course in quantum mechanics, for example, \cite{lan77}) that the transition dipole moment can be written as
\be
    \la d_j \ra = 
    -\frac{a_0}{Z} \int\limits_0^\infty \td{\psi}_{n_2 l_2 m_2}^\dagger(\td{\br}) e \td r_j \td{\psi}_{n_1 l_1 m_1}(\td{\br}) d^3 \td{r} = \frac{1}{Z} \la d^H_j\ra,
\ee
where $\td{r} = Z r/a_0$ is a dimensionless variable, 
$a_0 =\hbar^2 / m e^2 $ is the Bohr radius and $\td \psi_{n_i l_i m_i}$ are the wave functions
of  the electron in the Hydrogen-like atom written in terms of $\td r.$
The integral is independent of $Z.$ We observe that there is a simple scaling for the A-coefficients
between hydrogenic Z ion and H
\begin{equation}
	A_{n_1\to n_2}= Z^4\frac{4 (\w^H_{n_1\to n_2})^3}{3 \hbar c^3} (d^H_{n_1\to n_2})^2 
	= Z^4 A^H_{n_1\to n_2}.
\end{equation}

\subsection{Recombination Rate Coefficients}

Recombination coefficients and recombination cross sections for a free electron with the hydrogenic nucleus in its exact form cannot be simply scaled from HI. However, for our application, the recombining electrons are non-relativistic and we restrict our attention to this limit to obtain the scaling in the leading, dipole approximation. Assuming the recombining electrons are non-relativistic implies that the energy of the emitted photon is much less than the electron mass\footnote{In this section we work in the standard for QED units $\hbar = c = 1$ to avoid cluttering.} $\w \ll m.$ The recombination cross section can then be written as (see \cite{ber82} and references there in)
\begin{equation}\label{sph}
	d \sig_{rec} \simeq e^2 \frac{\w m}{\pi p} |{\bf e v}_{if}|^2 d\Omega,
\end{equation}
where $p$ is the momentum of the incoming electron, $\w$ is the energy of the outgoing photon, $\bf{ e}$ is the photon polarization vector, $d\W$ is the angular measure and ${\bf v}_{if}$ is the transition element
$ {\bf v}_{if}=\int \psi^\dag_f {\bf v} \psi_i d^3 x$ and $ {\bf v}=-\frac{i}{m}\nabla.$
Here $\psi_i$ and $\psi_f$ are the initial and the final wave functions of the electron. 

The initial electron wave function is the continuous spectrum wave function in the attractive potential of the hydrogenic nucleus $V=-Ze^2/r.$ For its explicit form see, for example, \cite{lan77}.
The final wave function of the electron is the discrete spectrum wave function in the attractive potential of the Z-ion nucleus, i.e.
 the electron wave function in the hydrogenic ion with Z charged nucleus discussed in the previous section.

One can show that the transition element can be written as
\be
    {\bf v}_{if}(p\to nlm)=
    -\frac{i}{m} \left(Ze^2 m\right)^{-1/2} \int \td {\psi}^\dag_f(\td {\bf r}) {\td {\nabla}} \td{\psi}_i (\td {\bf r}) d^3 \td{r}
    = Z^{-1/2}{\bf v}^H_{if}(\frac{p}{Z} \to nlm ).
\ee
Here
$\td{\psi}_i$ and $\td{\psi}_f$ are the initial and final wave functions written in terms of dimensionless variables $ \td{r} = Ze^2 m r$ and $\td{p} = \frac{p}{Ze^2m}, $ and $\td \nabla = \del/\del \td {\bf r}.$

Change of the momentum from $p$ for the hydrogenic ion to $p/Z$ for HI becomes obvious if we examine the energy conservation relation, 
 \be
    \size \no
    \frac{p^2}{2m} = E^Z_n + \w
	\quad {\rm vs.} \quad
   \frac{(p/Z)^2}{2m} = E^H_n + (\w/Z^2),
\ee
where $ E_n^Z = Z^2 E^H_n=- \frac{Z^2e^4 m}{2 n^2}$ is the energy of an electron on n's level of hydrogenic ions.
For the recombination cross section we find 
\be
\size \no
d \sig_{rec}(p \to nlm) = e^2 Z \frac{1}{Z}
 \frac{\w/Z^2 m}{ \pi p/Z}
  |{\bf e}\td{\bf v}^H_{if}|^2 d\Omega
 = d\sigma^H_{rec}(\frac{p}{Z} \to nlm)
\ee
After integration over angles $d \Omega$ and averaging over projections of the
orbital moment $m$ and the photon polarizations ${\bf e},$ we find  $\sigma_{rec}(p \to nl) = \sigma^H_{rec}({p}/{Z} \to nl ).$

Lastly, to obtain recombination coefficients $\a_{nl}(T),$ we need to average $u\sigma_{nl}$ over a Maxwellian distribution for the electrons,
\begin{equation}
	\a_{nl}(T) = \int\limits^\infty_0 u \sig_{nl}(p) f(u,T) du, \quad
	f(u,T)=\frac{4}{\sqrt{\pi}} \left( \frac{m}{2 k T} \right)^{3/2} u^2 e^{-\frac{mu^2}{kT}}, 
    \quad p = mu.
\end{equation}
Changing variables $T \to T^2/Z^2$ and $p \to p/Z$ in the integration, we find the scaling
\begin{equation}
	\a_{nl}(T) = Z \a^H_{nl}(T/Z^2).
\end{equation}

\section{Application to He$^{++}$}\label{appen_comp}

Here, we use the results from the Appendix A to partially explain the numerical results from \cite{sto95}. To this end 
we make a simplifying assumptions that the recombination line emissivities are dominated by
recombination rates of He$^{++}$ and H$^+$ (then scale this ratio by the energies of their respective photons). The cascade 
following recombination is determined largely by radiative decay as described by the A-coefficients and to a much less extent by collisions. 
Under this assumptions we can write the emissivities as linear combinations of the recombination rates:
\be
	\e (n_1 \to n_2) \simeq \hbar \w_{n_1 \to n_2} \frac{A_{n_1 \to n_2}}{\sum A_{n_1 \to all}} \sum\limits_{m \geq n_1} \a_m C_{m \to n_1}.
\ee
Coefficients $C_{m \to n}$ describe cascading down from $m$ to $n$ and are the functions of 
branching ratios. The radiative branching ratios will be the same for HeII as HI 
since all A-coefficients scale simply as Z$^4$  (\S \ref{acoef}). Therefore $C_{m \to n}$ are independent of nuclei charge Z.

\begin{figure}[ht]
\epsscale{0.7}  
\plotone{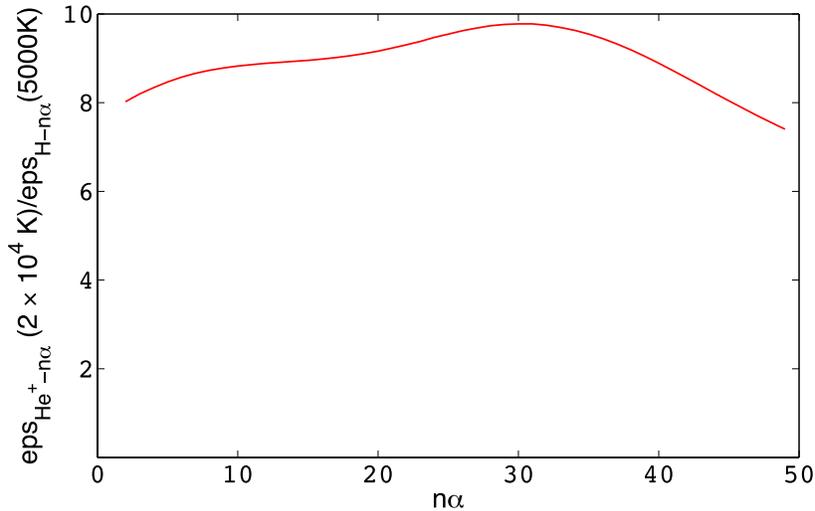}
\caption{The ratio of emissivities for HeII at 20000 K to HI at 5000 K from the full numerical results 
of \cite{sto95a} for comparison with the 'analytic' expectation of a ratio = 8.}
\label{exp2} 
\end{figure}

The recombination coefficients for HeII and HI, as derived in the Appendix A, scale as
\begin{equation} \label{ex1}
	\a^{\He^+}_{nl}(T) \simeq 2 \a^{\H}_{nl}(T/4).
\end{equation}
For the emissivity of $n\a$ line we obtain
\be\label{B2}
		\e_{\He^+-n\a}(T)\simeq 2^2   \hbar \w^H_{n+1 \to n}  \frac{A^H_{n+1 \to n}}{\sum A^H_{n+1 \to all}}  \times 2 
	\sum\limits_{m \geq n+1} \a^\H_m (T/4) C_{m \to n+1}
	\simeq 8 \e_{\H-n\a}(T/4).
\ee
In Figure \ref{exp2}  the emissivity ratios from \cite{sto95a} are shown for HeII at 20000 K compared to HI at 5000 K, illustrating that the numerical results reasonably confirm the approximate analytic prediction of a factor 8 difference. 

It is very hard and probably impossible to find exact scaling with temperature of recombination coefficients. So we use the result we found numerically above that
\be\label{B3}
   \frac{\e_{\H-n\a}(T_1)}{\e_{\H-n\a}(T_2)} \simeq \left( \frac{T_1}{T_2} \right)^{-4/3}.
\ee
Combining Equations \ref{B2} and \ref{B3}, we obtain the following relation between HeII and HI emissivities for the same n$\alpha$ lines:
\begin{equation}
	\frac{\e_{\He^+-n\a}(T)}{\e_{\H-n\a}(T)} \sim 
	8 \frac{\e_{\H-n\a}(T/4)}{\e_{\H-n\a}(T)}
	\sim 8 \, \left(\frac{1}{4}\right)^{-4/3}  \sim 50.
\end{equation}
The true numerical scaling factor varies between 50 and 65 for $n\a \leq 50 $ (see Fig. \ref{he_rel_h}), which reasonably confirms our  `analytical' prediction.

\vfill

\bibliography{scoville_recomb}

\end{document}